\documentclass[aps,prl,twocolumn]{revtex4}
\usepackage{graphicx}
\usepackage{revsymb}
\begin{document}

\noindent{\textbf{Comment on ``Solidification of a Supercooled Liquid
in a Narrow Channel''}}

In a recent paper \cite{ghomi}, Sabouri-Ghomi, et al. report on simulations
of solidification in a channel geometry using a phase-field method.  They
conducted two series of simulations, one with very small surface tension
(relative to the channel width), and the other at larger surface tension.
This system has been studied extensively both analytically and numerically 
\cite{kkl,bgt,bmkst,ihle,kkb-j,bb-a}
The authors do not relate to the more recent of these studies \cite{bmkst,ihle,kkb-j,bb-a} which in fact explain many of the results the authors claim 
as surprising and can correct some misinterpretations presented.

Let us address the large surface tension case first.  The authors find
a transition from a tip widening instability for low undercooling,
$\Delta \alt 0.5$, to
a stable finger for $\Delta \agt 0.5$.  The tip widening instability, as noted
by \cite{bmkst} is a result of the lack of steady-state solutions
for sufficiently low $\Delta$.  This threshold in $\Delta$ is a function
of surface tension $d_0$ and anisotropy $\epsilon$.  
Thus, for example, at $d_0 = 0.01$, (in units where the channel
width $L=2$) there are no steady-state
solutions below $\Delta \approx 0.62$ for zero 
anisotropy and below $\Delta \approx 0.60$ at $\epsilon=0.1$ \cite{kkb-j}. 
(Note that we define $\epsilon$ as the anisotropy of the
surface {\textit{tension}} and not of the surface {\textit{energy}}, as
in Ref. \cite{ghomi}, which
is a factor of 15 smaller for 4-fold anisotropy.) This threshold
decreases with $d_0$, going to 0 as $d_0 \to 0$ and also with
anisotropy.  
Above this threshold,
there exist a bilaterally symmetric finger, which is the channel
analogue of the free dendrite.  
We have extended the calculation \cite{kkb-j} of the threshold to the 
parameters of the large surface tension case of Ref. \cite{ghomi}, 
namely (in our
units) $d_0=0.00563$ and $\epsilon=0.75$.  The results 
are presented in Fig. 1, where we show the P\'eclet number, (dimensionless
velocity), $p$, as a function of $\Delta$. We see that the transition from
the unstable Saffman-Taylor branch to the dendritic branch occurs at
$\Delta=0.49$, in good agreement with the onset of stable dendritic
growth reported in Ref. \cite{ghomi}.  The stable pattern seen
above the threshold is thus nothing but the well-known channel analogue of the 
dendrite \cite{bmkst,kkb-j},
and not a new type of solution.  For example, we show in Fig. 2 the
dendrite for $\Delta=0.55$, which compares well with the pattern
observed by the authors at this undercooling (see their Fig. 4).
Also, this discussion should make clear that, contrary to the
authors' speculations, there is nothing
special happening at $\Delta=0.5$, which is of course the distinguished
 value in the Saffman-Taylor problem.
We should
also point out that at large $\Delta$, there also exist nonsymmetric 
fingers \cite{bmkst,ihle,kkb-j,bb-a},
which have been termed parity-breaking dendrites, or doublons.  The
authors appear to have imposed reflection symmetry about the midline
of their channel, which of course precludes seeing such solutions.

\begin{figure}
\includegraphics[width=2.5in]{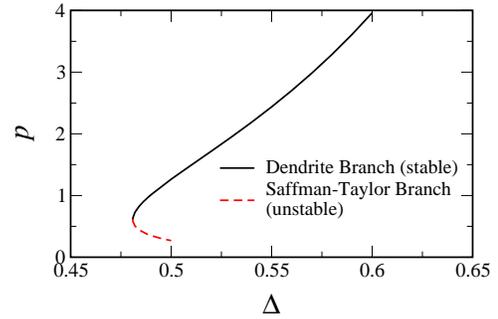}
\caption{P\'eclet number vs. supersaturation $\Delta$ for symmetric
solutions in a channel at $d_0=0.00563$ and $\epsilon=0.75$}
\end{figure}

\begin{figure}
\includegraphics[width=2.5in]{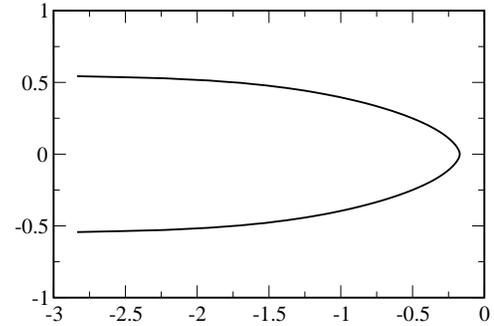}
\caption{Dendrite for $\Delta=0.55$ for parameters as in Fig. 1}
\end{figure}

For small $d_0$, the authors only find fingers which slow down in time,
and widen correspondingly.  The suprise here is that the authors did
not see stable translating dendrites.  However, these simulations were 
performed at very
small $\Delta$, so it is possible that again the authors are below the
threshold for dendrites.  We in fact suspect that the effective $d_0$ for
this series of runs is in fact much larger than the nominal value
$d_0=1.6\times 10^{-5}$ quoted, since it is extremely difficult to
adequately resolve the interface over such small scales, even with
adaptive gridding.

\leftline{Efim A. Brener}
\leftline{\quad Institut f\"ur Festk\"orperforschung, Forschungszentrum J\"ulich}
\leftline{\quad D-52425 J\"ulich, Germany}

\leftline{David A. Kessler}
\leftline{\quad Dept. of Physics, Bar-Ilan Univ.}
\leftline{\quad Ramat-Gan, IL52900 Israel}

\end{document}